\documentclass[12pt,preprint]{aastex}  
\usepackage{psfig}

\slugcomment{Version of \today}
\shortauthors{HENRY \& SMITH}
\shorttitle{ROTATION AND CYCLES IN $\gamma$ CAS} 
\begin{document} 

\newcommand{\gam}{$\gamma$\,Cas}

\title{ROTATIONAL AND CYCLICAL VARIABILITY \\ IN $\gamma$ CASSIOPEIAE. II.
FIFTEEN SEASONS}
 
\author{GREGORY W. HENRY} 
\affil{Center of Excellence in Information Systems,  \\
        Tennessee State University,  \\
	3500 John A. Merritt Blvd., Box 9501, Nashville, TN 37209;  \\
        gregory.w.henry@gmail.com}
\and

\author{MYRON A. SMITH} 
\affil{Catholic University of America,\\
        3700 San Martin Dr.,
        Baltimore, MD 21218;           \\
        msmith@stsci.edu}

\clearpage
\begin{abstract}

  The B0.5\,IVe star \gam\ is of great interest because it is the prototype 
of a small group of classical Be stars having hard X-ray emission of unknown 
origin.  We discuss results from ongoing $B$ and $V$ observations of 
the \gam\ star-disk system acquired with an Automated Photometric 
Telescope during the observing seasons 1997--2011.  In an earlier study, 
Smith, Henry, \& Vishniac showed that light variations in \gam\ are 
dominated by a series of comparatively prominent cycles with amplitudes of 
0.02--0.03 mag and lengths of 2--3 months, superimposed on a 1.21-day 
periodic signal some five times smaller, which they attributed to rotation.  
The cycle lengths clustered around 70 days, with a total range of 50--91 
days. Changes in both cycle length and amplitude were observed from year to 
year.  These authors also found the $V$-band cycles to be 30--40\% larger than 
the $B$-band cycles.  In the present study we find continued evidence for
these variability patterns and for the bimodal distribution of the 
$\Delta$$B$/$\Delta$$V$ amplitude ratios in the long cycles. During the 2010 
observing season, \gam\ underwent a mass loss event (``outburst''), as 
evidenced by the brightening and reddening seen in our new photometry.  
This episode coincided with a waning of the amplitude in the ongoing cycle.  
The Be outburst ended the following year, and the light-curve amplitude 
returned to pre-outburst levels.  This behavior reinforces the interpretation 
that cycles arise from a global disk instability.  We have determined a more 
precise value of the rotation period, 1.215811${\pm 0.000030}$ days, using 
the longer 15-season dataset and combining solutions from the $V$ and $B$ 
light curves.  Remarkably, we also find that both the amplitude and the 
asymmetry of the rotational waveform changed over the years.  
We review arguments for this modulation arising from transits of a surface 
magnetic disturbance.  Finally, to a limit of 5 mmag, we find no evidence 
for any photometric variation corresponding to the \gam\ binary period, 
203.55 days, or to the first few harmonics.

\end{abstract}
 

\clearpage

\section{Introduction}
\label{intro}

  \gam\ (HD\,5394, B0.5\,IVe) is the prototype of the classical Be stars,
defined as non-supergiant, non-pre-main-sequence B stars with Balmer-line 
emission recorded sometime in their history but without ongoing mass infall.
The Balmer-line emission arises from a flattened Keplerian disk.  
Be stars have broad photospheric lines due to their rapid rotation 
(e.g., Porter \& Rivinius 2003).  \gam\ was the first star with H$\alpha$ 
emission to be reported (Secchi 1866). It became important again with the
discovery of its copious hard X-ray emission (Jernigan 1976), which is 
unusual for early-type stars (White et al. 1982; G\"udel \& Naz\'e 2009). 
Its high-energy 
peculiarities include an X-ray light curve that varies on several 
timescales ranging from seconds to 2--3 months (Smith, Robinson, \& 
Corbet 1998, hereafter SRC98; Robinson, Smith, \& Henry 2002, hereafter 
``RSH"), a luminosity L$_{X_{2-10}}$ $\sim$  10$^{33}$ ergs\,s$^{-1}$, and an 
X-ray spectrum that reveals the presence of 3--4 optically thin hot plasma 
components.  \gam\ is now one of a group of at least nine recognized Galactic 
X-ray emitting B0.5e--B1.5e stars (Lopes de Oliveira et al. 2006; Smith et al. 
2012, hereafter S12).  The source of the X-rays remains a source of spirited 
debate.  The two most common suggestions are that the emission comes from 
accretion from the Be star wind or disk (White et al. 1982) or from the
interaction of magnetic field lines from the star and its decretion 
(mass loss) disk, with an ensuing release of high-velocity particle
streams impacting the Be star and causing explosive high-energy events, 
often dubbed ``flares" (see Robinson \& Smith 2000, hereafter RS). 
In the latter scenario, flares are not necessarily products of local 
magnetic instabilities, such as solar-type flares.  The difficulty in 
settling the controversy comes from the fact that the X-rays can arise,
in principal, from any environment in which particles (electrons or 
protons) can be accelerated to high velocities, either by release of 
tension in stretched magnetic field lines or in a steep gravitational 
potential.  Probably the most telling piece of evidence lies in the 
simultaneous rapid variations of X-ray flux and UV and/or optical 
signatures that are presumed to form close to the Be star (e.g., 
Slettebak \& Snow 1978; Peters 1982).  To get a handle on this, SRC98
launched a 21-hour campaign on \gam\ that included X-ray observations by 
the {\it Rossi X-ray Timing Explorer (RXTE)} and spectroscopic and 
photometric observations by the {\it Hubble Space Telescope (HST)}.  The 
latter component was a very high time-resolution monitoring of the 
Si\,IV 1394--1403\,\AA\ resonance line doublet and a surrounding quasi 
continuum.  The latter could be binned in wavelength to give a time 
record of both the UV continuum and several lines that were formed in 
the photosphere.  The UV continuum record showed a pair of faint dips that 
were investigated by Smith, Robinson\, \& Hatzes (1998), who demonstrated that 
these absorption features arose from cool, translucent clouds that transit 
the star.  These structures transited too rapidly to be surface features, 
nor could they easily be due to structures within the Keplerian disk
because Long Baseline Optical Imaging has demonstrated that the disk plane 
is tilted by about 48${\pm 4}$$^\circ$ with respect to our line of sight to
the star (e.g., Stee et al. 2012). These same authors noted similar dip 
features several days later in spectra from {\it International Ultraviolet 
Explorer (IUE)}.  The UV spectra evinced narrow components moving 
blue-to-red across the line profiles, consistent with optical spectra of 
other observers (e.g., Yang, Ninkov, \& Walker 1988; Smith 1995).  These 
so-called {\it migrating subfeatures (msf)} have also been observed in a 
He\,I line of another \gam\ star, HD\,110432 (Smith \& Balona 1998). Such
features have long been observed in the magnetically active pre-main-sequence 
K5e star, AB\,Dor (Cameron-Collier \& Robinson 1989).

  \gam\ is now known to be the primary star of a wide binary system with 
P = $203.5 \pm 0.20$ days (Harmanec et al. 2000; Miroschnichenko et al. 
2001; Nemravov\'a et al. 2012; S12). The last three of the 
cited studies agree that the binary system is nearly circular ($e<0.03$), 
and the unseen secondary has a mass of ${\frac 12}$--1 M$_{\odot}$.  
At least two ``\gam\ analogs" are blue stragglers, according to 
their likely memberships in galactic clusters (e.g., Marco et al. 2009).  
Conceivably all of them are blue stragglers, although this is not yet well
established. Moreover, no evidence exists for \gam\ being a blue straggler 
since it is not known to be associated with a galactic cluster.  
Aside from the obvious possibility 
that the secondaries of such stars could accrete matter from the Be star, 
it is also possible that their previous evolutionary history could 
have resulted in mass and angular momentum transfer to the current primary, 
thereby spinning it up.  The fact that \gam\ rotates close to its critical 
velocity may be part of what is needed for the X-ray generation mechanism 
to develop. 

  Although \gam\ has been observed by nearly all X-ray space telescopes, its 
brightness ($V$ $\approx$ 2.1) makes it inaccessible to most modern UV and 
optical instruments because it saturates their detectors.  However, with 
suitable use of neutral density filters, \gam\ has provided an excellent 
target for monitoring with the T3 0.4m Automatic Photometric Telescope (APT) 
in Arizona (see \S2 below).  Ironically, long-term optical monitoring by 
this small, ground-based telescope has resulted in a series of major 
discoveries, some or all of which may turn out to have bearing on the X-ray 
emission mechanism, which orbiting X-ray telescopes themselves could not 
elucidate. These include:

\begin{itemize}

\item the presence of a robust period of 1.21 days in optical flux (Smith, 
Henry, \& Vishniac 2006, hereafter SHV) with an unusual waveform.  There 
is no detectable variation in the $B-V$ color index at this period, 
suggesting the variation is due to rotational advection of a surface feature.
We will address the rotational interpretation of this feature in
$\S$\ref{rotmod}.

\item the existence of long optical cycles of 50--91 days, 
which seem to correlate with corresponding X-ray cycles (RSH; SHV). 
These short-lived cycles have larger amplitudes in the Johnson $V$ band 
than in the $B$ band. Because the variations are ``red," they probably 
arise in the Be disk, which contributes 10--50\% of the 
system's integrated light at visible wavelengths. This percentage increases 
rapidly in the infrared, where at least the inner disk becomes optically 
thick because of hydrogen's dominant bound-free and free-free transitions.

\item a brightening and reddening of the star-disk system in 2010, 
reaching a peak in early 2011, then fading through the end of our dataset 
(2012 February 24).  There was a corresponding increase in H$\alpha$ 
emission (S12), and we refer to this extended event as the ``2010 outburst."  
The outburst was also correlated with the second and longest of a two-density 
column system observed in contemporaneous (2010 July--August) high-resolution 
X-ray spectra. The increase in this column is manifested by a decrease 
in soft X-ray band flux (${\frac 12}$--1 keV) relative to flux in the 
hard band (2--8 keV), as in all previous epochs for which these fluxes
are recorded (S12). 

\end{itemize}

  This paper presents the results of our ongoing photometric monitoring 
campaign on \gam. SHV covered observing Seasons 1997--2004. This paper 
extends the time coverage through the 2011 observing season.\footnote{\gam\ 
comes to opposition in early October, and we define the \gam\ observing 
seasons, e.g., the ``2004'' observing season, as the time interval 2004 
May--2005 February, minus the months of July and August lost to the monsoon 
rains that come to southern Arizona every summer.}  Our observations 
include both the 2010 outburst and also the next season during which the
brightness declined, though not yet to its level at the 
start of the outburst.  We analyze the long cycles in the new seasons, 
including their color variations, and search for additional long cycles. We also
reanalyze the long-cycle subtracted residuals to better define the 1.21 day 
variation and its behavior with time.

\section{The Automated Photometric Telescope Observations}
\label{obsn}

  Our optical continuum observations of \gam\ were acquired with the T3 
0.4m Automated Photometric Telescope (APT) at Fairborn Observatory, located 
in the Patagonia Mountains of southern Arizona.  T3 is one of eight 
Tennessee State University (TSU) telescopes operated at Fairborn for 
automated photometry, imaging, and spectroscopy (Eaton, Henry, \& Fekel 2003; 
Eaton \& Williamson 2007).  

  Full details concerning our \gam\ observing program can be found in 
RSH and SHV, and details of telescope operations and data 
reduction procedures can be found in Henry (1995a, 1995b).  Briefly, 
the T3 APT was programmed to observe \gam\ once every two hours on 
every clear night throughout its observing season.  For each 
observation, the APT cycles between \gam\ and the nearby comparison and 
check stars HD\,6210 (F6\,V) and HD\,5395 (G8\,IIIb), respectively, in the 
following sequence, termed a group observation: K, S, C, V, C, V, C, V, C, S, 
K, where K \& C are the check and comparison stars, V is \gam\,, and S is 
a sky reading.  Each \gam\ measurement is thus bracketed by the comparison 
star on both sides three times during each group observation.  All 
observations were made in the standard Johnson $B$ and $V$ passbands 
with 10\,s integrations for \gam\ and HD~5395 and 20\,s for HD 6210 and 
the sky readings.  In addition, all measurements (except for the 1997
observing season) were made through a 3.8~mag neutral density filter 
to attenuate the counts from \gam\ and so avoid large deadtime 
corrections. However, as noted in SHV, a different neutral density 
filter was used for the 1997 season from the more permanent one we 
selected in 1998, and thus the 1997 season has not been used in those 
analyses requiring linkage among the seasons.  A complete sequence, 
including both $B$ and $V$ filters, requires about 8 minutes to 
execute.  Group mean $V-C$ and $K-C$ differential magnitudes are computed 
and standardized with nightly extinction and yearly-mean transformation 
coefficients determined from nightly observations of a network of standard 
stars.  Group means are taken to be single observations thereafter. Typical 
{\it rms} errors for a single observation, as measured from pairs of constant 
stars, are $\pm{0.003-0.004}$~mag.  The 15 yearly-mean $K-C$ differential 
magnitudes scatter about their grand mean with an {\it rms} of only 
0.0019~mag, indicating very good long-term stability in the brightness of 
both the comparison and check stars as well as good stability of our ND
filters and calibrations to the Johnson photometric system.  

  The $V-C$ differential magnitudes were converted to apparent magnitudes
of \gam\ by assuming apparent magnitudes m$_V$ = 5.84 and m$_B$ = 6.40 
for the comparison star.  Aside from our inability to properly calibrate 
the 1997 season with the data from subsequent seasons, our seasonal-mean 
brightness values should have an internal {\it precision} of approximately 
${\pm 0.002}$ mag, as shown by the seasonal mean $K-C$ values, while the ND
filter limits the {\it accuracy} of our apparent magnitudes to 1--2\%. 

  SHV presented 3135 $V$ and 3157 $B$ observations covering the nine 
observing Seasons 1997--2005. In 1998 the filter was changed to another, 
which has been used in the meantime.  Here, we present data from six more 
observing seasons 2006--2011, for a total of 4675 $V$ and 4716 $B$ 
observations.  All 15 years of data (1997--2011) are listed in Table\,1, 
available in the online version of this paper. 

\vspace*{.15in}
\begin{table}[!h]    
\tablenum{1}
\begin{center}
$Table~1:$
Photometric Observations of \gam\ in Seasons 2006-2011
\begin{tabular}{c|c|c|c|c}
\tableline\tableline
Reduced Julian Date & Var $B$ & Var $V$ & Chk $B$ & Chk $V$ \\
RJD                 & (mag)  & (mag)  &  (mag)  &  (mag) \\
\tableline
50718.6966 & $-$4.393 & $-$3.641 & $-$0.818 & $-$1.216 \\
50718.9253 & $-$4.393 & $-$3.642 & $-$0.814 & $-$1.222 \\
50720.7936 & $-$4.386 & $-$3.644 & $-$0.821 & $-$1.218 \\
50720.9191 & $-$4.387 & $-$3.638 &   99.999 & $-$1.223 \\
50721.6940 &   99.999 & $-$3.648 &   99.999 &   99.999 \\
50721.7933 & $-$4.386 & $-$3.639 & $-$0.816 & $-$1.212 \\
\tableline
\end{tabular}
\end{center}
\tablecomments{Table 1 is presented in its entirety in the electronic
edition of the {\it Astrophysical Journal.}  A portion is shown here
for guidance in data format and content.}
\tablecomments{An entry of 99.999 signifies that the differential magnitude 
was discarded because its internal standard deviation exceeded 0.01 mag, 
indicating nonphotometric conditions.}
\end{table}
\vspace*{.15in}

\section{RESULTS}
\subsection{Long-Term Brightness and Color Variations}
\label{yr2yr}

\begin{figure}[h!]
\centerline{
\includegraphics[scale=.35,angle=90]
{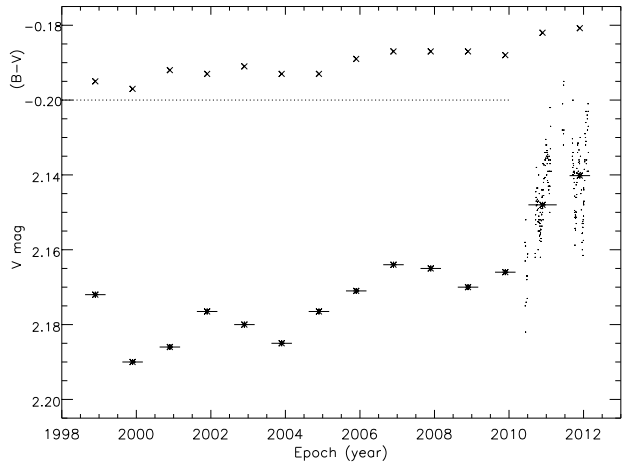}}
\caption{ 
Johnson $V$ (m$_V$) magnitudes ($\ast$ symbols) and $(B-V)$ color 
indices (crosses) of \gam\ acquired with the T3 APT at Fairborn 
Observatory during the 1998--2011 observing seasons; large symbols 
represent seasonal means.  Error bars on the means are smaller than 
the symbol size.  Horizontal bars denote the coverage for a seasonal mean
time.  Small dots indicate individual m$_V$ measurements 
in the 2010 and 2011 seasons.  All magnitudes in this paper are 
plotted with brightness increasing upward.
}
\label{vbv}
\end{figure}

  In Figure\,\ref{vbv}, we present the seasonal mean $V$ magnitudes and 
$B-V$ color indices for the 1998--2011 observing seasons.  The figure shows 
that the light from the Be star-disk system brightened and reddened 
during most of this interval, in fact by a total of about 5\%.
As noted by S12, half of the brightening and reddening increases 
started in mid-2010, fortuitously
coinciding with a simultaneous campaign of {\it XMM-Newton} and Long 
Baseline Optical Interferometric observations. This event, an example 
of what is referred to as an ``outburst" of a classical Be star,
reached peak brightness shortly before APT observations resumed 
on 2011 June 12, the last observing season analyzed in this paper.

\begin{figure}[h!]
\centerline{
\includegraphics[scale=.35,angle=90]
{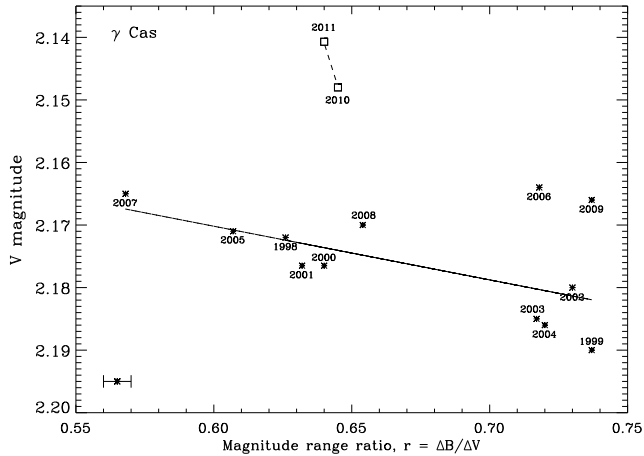}}
\caption{ 
The seasonal quantities $r$ = $\Delta$$B$/$\Delta$$V$ discussed in
the text plotted against seasonal $V$ means for \gam.\ For each of the
two filters, the ``$\Delta$" quantity is the mean of the absolute values 
of all observations in a season with respect to the seasonal average.  
Season annotations are indicated near each plotted point.  The solid line 
is the regression fit through the points, except for of the 2010 outburst 
\& 2011 decline intervals. 
}
\label{deltr} 
\end{figure}

  Although brightness variations in $B$ and $V$ are highly correlated, the 
flux recorded in the $V$ band is more sensitive to brightening events than 
$B$, implying the activity is concentrated in the disk. Following SHV, we 
parameterize this by computing for each season a magnitude range ratio 
$r$ = $\Delta B$/$\Delta V$, where $\Delta B$ and $\Delta V$ are computed 
as the means of the differences of the absolute values of each observation 
in a season with respect to the seasonal mean values, $<B>$ and $<V>$.  
So defined, the $r$ index tends to unity for variations 
in the photosphere and to values of 0.5--0.7 for structural changes in the 
disks of Be stars in general and of \gam\ in particular (Harmanec et 
al. 1980; Hirata 1982; Barylak \& Doazan 1986; Hirata 1995).  
Figure\,\ref{deltr} exhibits the regression line formed from the 1998--2009 
seasons.  The correlation of $r$ with system brightness $V$ is marginally 
significant (at 2.5$\sigma$) and would be lower with the addition of the 
last two seasons.  In Fig.\,\ref{deltr} the symbols for these two seasons 
are connected by a dashed line, indicating the $r/V$ trajectory during the 
outburst.  The brightening observed during this period can be interpreted as 
a simultaneous injection of mass from the star to the disk. (Of course
the attenuation of the soft X-rays indicates that the expelled gas also
occupies a larger volume that includes our sight line to the star.) 
Within the errors, neither the $r$ value nor the period of the long cycle 
seems to have been affected by the outburst. 

\begin{figure}[h!]
\centerline{
\includegraphics[scale=.35,angle=90]
{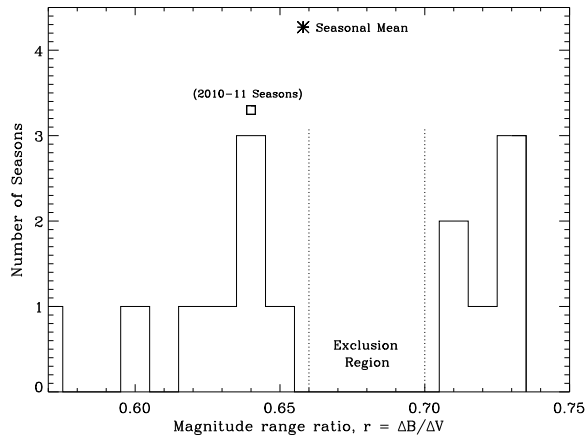}}
\caption{ 
Histogram of the r values (defined as in Fig.\,\ref{deltr}).  To a 
modest level of significance the distribution is bimodal.
The mean $r$ for all seasons, except the anomalous Seasons 2010 and 2011, 
is shown by an asterisk. The mean value of the last two seasons is 
labeled separately.
}
\label{histr}
\end{figure}

  In their study of light curve averages for seasons 1997--2005, SHV 
tentatively concluded that the values of the seasonal $r$ ratios tend to 
clump around two values, near 0.62 and near 0.72, with a gap between them.  
The additional seasons in this study, shown in Figure\,\ref{histr}, tend
to confirm the impression of two separated groups. 
A Monte Carlo analysis of an assumed distribution of these seasonal 
amplitude ratios shows that the gap is marginally significant (to 2$\sigma$).
In this exercise our bin sizes, were 0.01 magnitudes, which is closely
matched to the {\it rms} of ${\pm 0.009}$ measured from the data
subsets from which the r values were computed.
This implies that the amplitude ratio $r$ has a weak dependence on its 
recent history -- that is, what may be important is the question: has the 
disk been bright or faint in recent times?

\subsection{The Long-Term Cycles}
\label{lng}

   RSH and SHV found that the long cycles of $\sim$70 days are the 
dominant source of brightness variation in the optical light curve of 
\gam.~  For some seasons SHV were able to fit these variations well 
in both $B$ and $V$ with a single sinusoid.  In these cases the 
brightness variation could be fit with four parameters:  slope, period 
(cycle length), amplitude, and mean brightness level of the sinusoid.  
However, for most seasons in the current study, we had to adopt a search 
for {\it ``am/fm"}-type modulations, that is, we were usually forced to 
consider linear changes with time in the amplitude {\it (am)} and/or 
length {\it (fm)} of our sinusoids to obtain good fits. Often a mild 
linear trend had to be imposed throughout a whole season.  The data were 
uncorrected for the 1.21-day rotational period discussed in the next 
section. This signal is typically five and sometimes ten times smaller 
than the cycle amplitudes.

  We used the {\it Interactive Data Language} to develop customized 
procedures for periodogram fitting the long cycles to the observations in 
each observing season.  We determined ``best fits" to the slope, cycle 
length, amplitude, and mean brightness with the MPFIT package published by 
Markwardt (2011).  However, a simple sinusoid-fitting procedure gave a
reasonable solution for only about half the seasons.  As noted by RSH and
SHV, the long cycles exhibit changes in all four parameters.  Sometimes it
was possible to link one season to the next with only small changes to these 
parameters, e.g., the 2006 \& 2007 and also the 2010 \& 2011 seasons. In
these two cases, only small changes in the slope and cycle length parameters, 
(within the formal uncertainties of the Markwarth solution) were needed to
link the two seasons. In other cases, it is debatable how the cycle lengths 
and amplitudes change with time. We also note that the time interval 
between seasons is 90${\pm 10}$ days. Therefore, we discarded this simple 
procedure in deference to economy of assumptions.  In most cases, our next 
step was to seek a solution linking roughly the later two thirds of one 
season with the initial two thirds of the next. Sometimes, even this 
procedure failed to give satisfactory fits because, as found by SHV for 
Seasons 1998 and 2003, a particular cycle could damp out in the interim.  
The cycle usually recovered within weeks but sometimes with different 
parameters.  SHV reported that in Season 2003 this occurred with a 
recovery of the former cycle's phase.  Because of these characteristics, 
we minimized the number of parameters by requiring that a new oscillation 
grow from an old one with the same phase at the null point.  Given the 
variety of behaviors, it is impossible to claim that the fits are unique 
because their qualitative characteristics can be quite different.  However, 
we are confident that alternative solutions are either worse or require more 
free parameters.  By fitting the $B$-band data over the same timespan as
the $V$ solutions and comparing best-fit parameters, we estimate typical 
errors in the cycle lengths to be about ${\pm 1}$ day, while the amplitudes
are estimated to be uncertain by 10--15\%. These estimates must 
be tempered by the uncertainty of whether or not the long-term trends in
the light curves vary linearly, i.e., free of other system noise sources, 
such as irregular injections of mass into the disk.

\begin{figure*}[ht!]
\centerline{
\includegraphics[scale=.70,angle=90]
{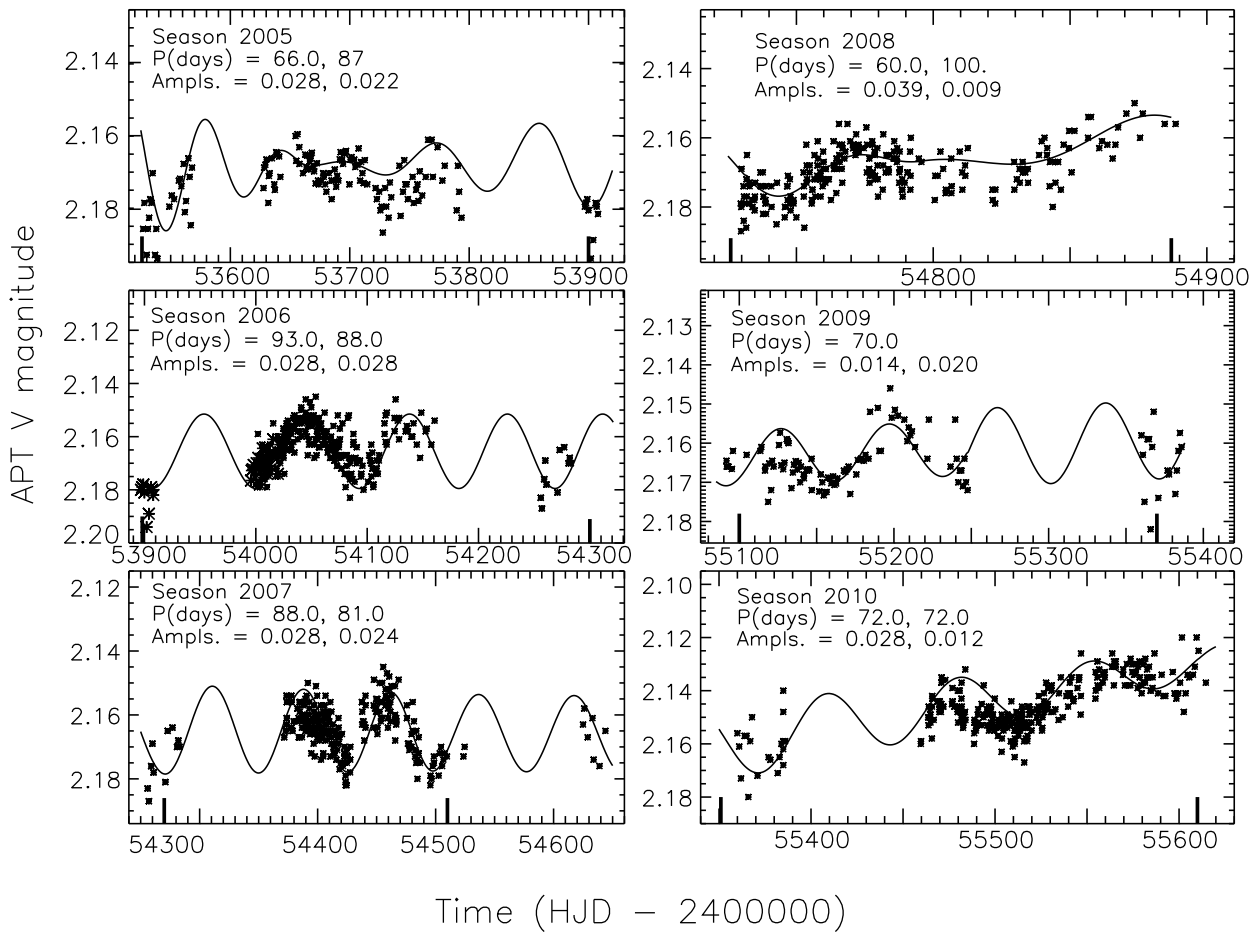}}
\caption{
$V$-band photometry of \gam\ showing the light curve during Seasons 2005--2010,
ending in 2011 February.  The approximate date of the immediate post-maximum 
of 2010--2011 is indicated by a small clump of points at 2011.5.  The solid 
line indicates the damped or changing sinusoids needed to fit the data.
Their combs at the bottom of the panels show the reference start and end times
for the computation of these sinusoids using the periods and amplitudes
indicated.
}
\label{seas0813}
\end{figure*}

\begin{figure}[h!]
\centerline{
\includegraphics[scale=.35,angle=90]
{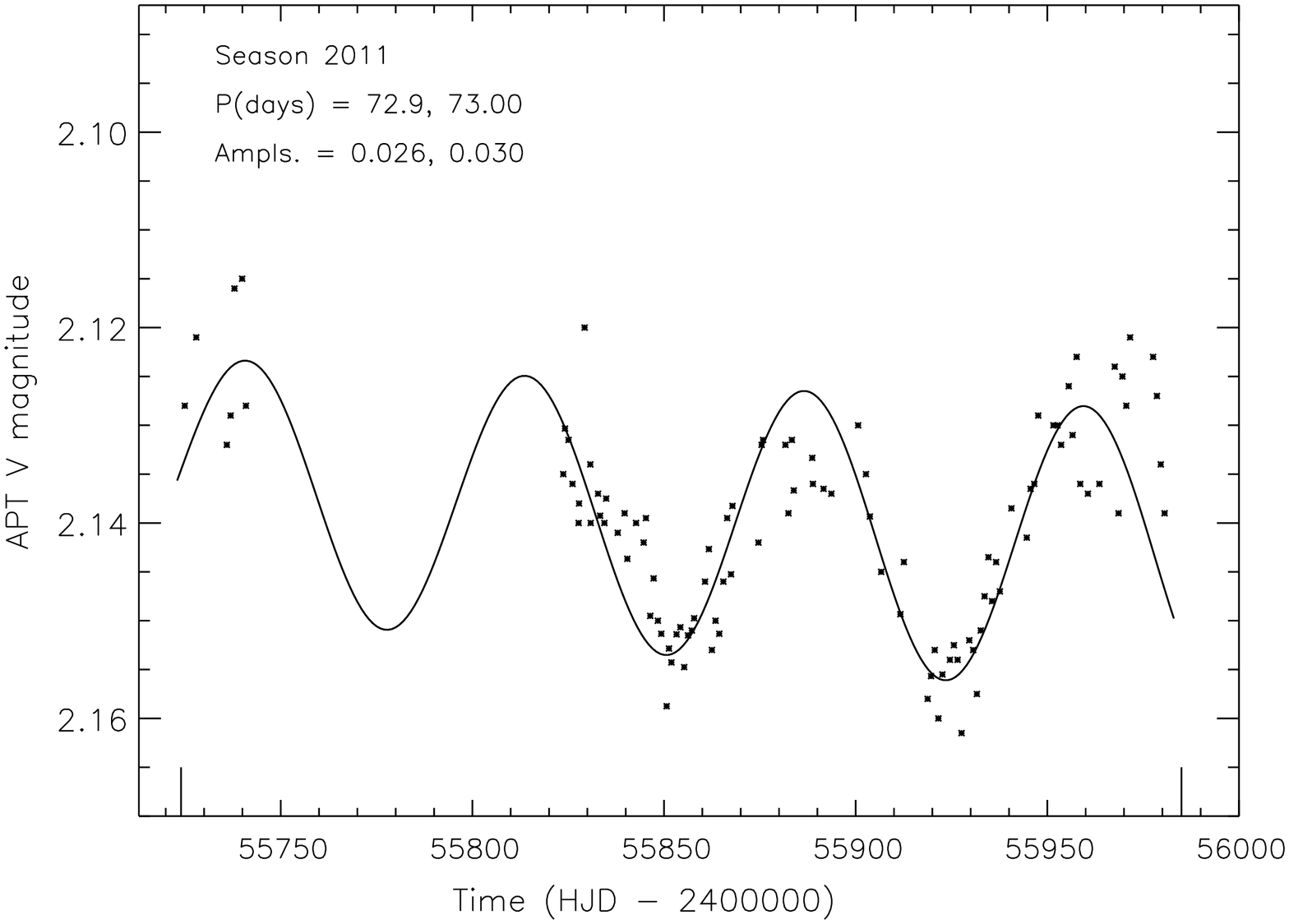}}
\caption{
$V$-band photometry of \gam\ showing the long cycles of Season 2011.
Note the trend to fainter magnitudes after reaching a maximum in 
mid-2011 following the 2010 ``outburst."
}
\label{seas14}
\end{figure}

  Figures \ref{seas0813} and \ref{seas14} depict our best-fit solutions to 
the nightly means of Seasons 2005--2011.  Each panel in Fig. \ref{seas0813} 
shows the solution for each season changing smoothly into the solution of 
the next.  Therefore, data in the middle of a particular season reflect a 
continuous solution from the previous to the next season. The quality of 
the solutions is similar to those discovered by SHV. 

  The Season 2006 data suggest a cycle length of $\sim$93 days, while 
Season 2008 suggests the growth of a new, rather long cycle, {\it perhaps} 
$\sim$100 days in length.  These cycle lengths are slightly longer than 
the 91 day length that SHV encountered for Season 2000.  The variations 
in Season 2007 alone suggest a solution that can be extrapolated to much 
of the next season.  Even so, the brightness during the beginning of this 
season (RJD = 54260) suggests that the star-disk system was abnormally faint.  
In particular, Season 2005 exhibits a damped oscillation that grew to a new 
cycle in 2006, with no evident change in phase. In the first part of Season 
2009 a new cycle has started from the previous season. In fact, we suspect 
that the matter injection into the disk was not regular either. This would 
account for the unusually poor fit at this time.  From the inadequate fit at 
RJD = 55100--55150 in Season 2009, we suspect that a new cycle started 
growing in this interval. Unfortunately, inclement weather in 2009 June 
meant we could not determine what the trend in the variations were at this 
critical time.  For the other seasons, the complexity of the parameterization 
invoked for the fits is not a fault of inadequate time coverage or noisy data.  This can be verified by independent solutions for the $B$ band, which show 
similar details in the departures from the data. 

  The two least satisfactory fits were during times of anomalous behavior 
of the light curve in Seasons 2005 and 2010 --- see Fig.\,\ref{seas0813}.
The first of these was during yet another damping of an old cycle (period) 
and growth of a new one. The brightness decreased during this interval.
The other inadequate fit occurs in the middle of Season 2010 during the
growth of the 2010 outburst. Again, this flaw may be 
related to the photometric/spectroscopic outburst during this time.  
We also notice from Fig.\,\ref{seas14} that not only did the cycle persist 
but its amplitude grew to a more typical value after the event ceased. 
The restored amplitude is sufficiently large that the oscillation can be 
discerned in the $B-V$ color index during this interval.

  For the 2005--2011 observing seasons analyzed in this paper
all cycle lengths fall within the same range of 50--93 days observed by 
SHV for Seasons 1997--2004, where again the upper limit may be 100 days. 
The means and {\it rms} values of the cycle lengths are 76$\pm{11}$ day. 
Fully 20\% of the cycle lengths deviate by $\ge$20\% from this mean. 
Like SHV, we find the cycle amplitudes in the $V$ band photometry to 
be 0.006--0.030 magnitudes.

\subsection{The 1.21-day periodicity}
\label{rotper}

  The extraction of the 1.21 day signal found by SHV requires some caution,
first because the period is near one day and second because the amplitude is
nearly an order of magnitude smaller than the dominant long-cycle variation, 
which, according to SHV, is nonsinusoidal. We began by detrending each 
observing season of the $V$ and $B$ datasets and forcing each to a common 
mean brightness. We ignored obvious low-frequency peaks caused by the long 
cycles, and this allowed us to isolate a smaller peak at 0.8225 day$^{-1}$ 
that was found and identified as a rotational signature. In the time domain, 
this resulted in a predicted sinusoidal variation for the entire dataset 
that we refer to as the multiseasonal average. 

  SHV noted the possibility that the seasonal mean amplitudes and even
phases of this periodicity might vary, but they did not pursue the question 
further.  With the extended timeline of six more seasons, we decided to 
examine this possibility by imposing the ephemeris (period and phase of 
mininum light for the multiseason solution) and searching for variations 
in the phases and amplitudes in the individual seasonal datasets.  In fact,
we found substantial variations in both amplitude and reference phase from
season to season. The full amplitudes ranged from 4--8 mmag prior to 2004 
but ranged only between 1--3 mmag from 2005 onward. 

\begin{figure}[h!]
\centerline{
\includegraphics[scale=.35,angle=90]
{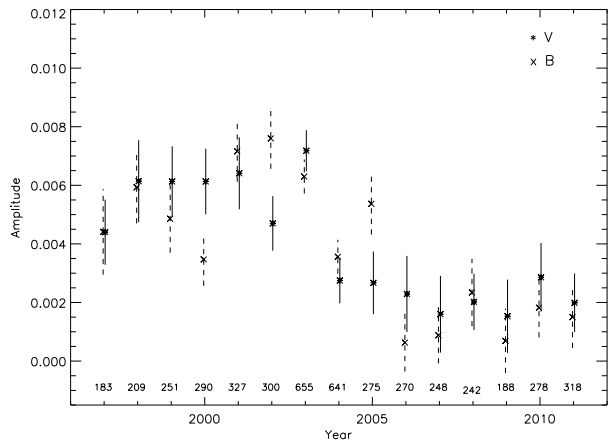}}
\caption{
Full amplitudes of the rotational signature in $B$ and $V$ for each of 
the 15 observing seasons. These values were determined by rectifying the 
seasonal means to a common value then prewhitening the 15-year $B$ and $V$ 
datasets for several periods consistent with the long cycles found in
various individual seasons. The numbers of $V$ observations in each 
season are given at the bottom of this and the following figure. 
}
\label{sf3}
\end{figure}

  To see whether these amplitude changes 
are caused by subtle effects from the dominant, long-period signals, we 
prewhitened the $V$ and $B$ band light curves by subtracting out several 
dominant peaks that correspond to the long cycle periods, using the method 
of Vani\^cek (1971), as implemented in SHV.  We then used least squares 
to find the best fit sinusoid in the residuals of the complete dataset for 
each filter (excluding Season 1997, with its uncertain zero points). Once 
the pair of rotational ephemerides were redetermined, we imposed them on 
the seasonal light curves and solved again for the mean amplitude and the 
deviation of the phase of minimum light as before. This time we found a 
more uniform signal with much less rapid (season to season) variations in the 
rotational amplitudes. However, the general trend in the amplitude variations 
remained the same. Figure\,\ref{sf3} depicts the behavior of the amplitudes, 
starting at a level of 5--7\,mmag in the early seasons, decreasing to 
1--2\,mmag from 2003 to 2005, and remaining at the lower level through the 
2011 season.

  Next, recognizing that sine curves do not always give a good representation 
of the long cycles, we subtracted the long-cycle fits shown in SHV and in our 
Figs.\,\ref{seas0813} and \ref{seas14} and repeated our analysis yet again.
We imposed an artificial sinusoid to the full $B$ and $V$ 
datasets with an amplitude of 3.0\,mmag and an arbitrary control period of 
1.3418644 days in place of the rotation signal.  We then attempted recovery 
of this period by fitting the light curve of each season.  Dividing the 
entire observing timeline into two equal segments of seven seasons 
(1997--2003 and 2005--2011; the omission of Season 2004 recognizes that this
season seems to be transitional), we found mean amplitudes in $V$ and $B$ 
of 3.77 and 3.90 ($\pm{1.5}$ mmag), respectively, for the first half and 
3.20 and 3.24 ($\pm{0.7}$ mmag) for the second.  We consider these slight 
differences to be a measure of the uncertainties introduced by our processing, 
incomplete subtractions of long-cycle signals, and periodogram sidelobes 
from gaps in our observing windows. 

\begin{figure}[ht!]
\centerline{
\includegraphics[scale=.35,angle=90]
{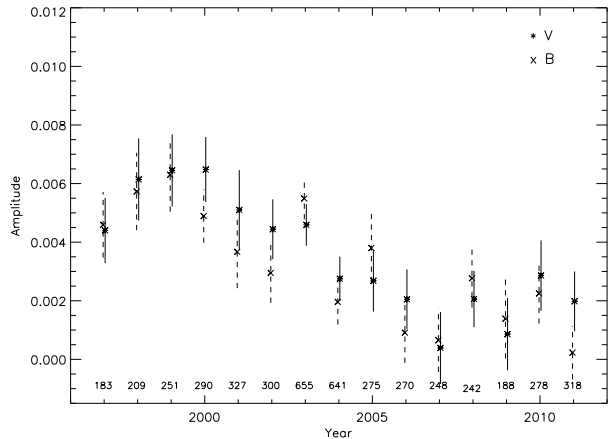}}
\caption{
Full amplitudes of the rotational signature in $B$ and $V$ for the 15 
individual seasons in our study. These values were determined from the 
light-curve residuals after removing the long-cycle modulations found in 
SHV and in Fig.\,\ref{seas0813}. 
}
\label{sf4}
\end{figure}

  In contrast to our experiment with a control period, results from the 
real-data residuals, shown in Figure\,\ref{sf4}, differ rather little from 
those of the previous analysis of real data and show a marked contrast to the
behavior of the control set. The primary difference is that the rotational
amplitudes decrease smoothly during Seasons 2002--2005.  For the first 
seven seasons, we found amplitudes averaging 5.4 and 4.8 (${\pm 1.05}$\,mmag) 
in $V$ and $B$, respectively, and for the last seven seasons 2.0 and 
1.8 (${\pm 1.01}$\,mmag), mean differences that are almost six times larger 
than the controls.  Moreover, the signals in the last several seasons are 
nearly undetectable, leading us to realize that had we started our program 
a few years later, when the amplitudes had already decreased, an important 
facet of the $\gamma$\,Cas puzzle -- the discovery of the rotational 
modulation -- might not have been unraveled for some time yet to come.  

  One challenge to this result might be that visible changes in the observed 
magnitudes that are caused by minor mass-loss episodes degrade the fit of our 
models, which are based on simple analytical functions.  For example, if the 
star were more active in ejecting mass during the last seven seasons, it 
would brighten the star in ways that we cannot anticipate.  This would 
degrade the fits to low-amplitude signals such as the rotational modulation, 
leading to lower observed amplitudes.  To check for this possibility, we 
computed the mean and {\it rms} of the systematic differences (absolute 
values) for all seasons of both the $B$ and $V$ light curves. The results 
showed these metrics to be constant over time, with no first/second half 
trends apparent. Thus, there is no evidence that the long cycles introduce 
long-term biases in our determination of the amplitudes of the 1.21-day 
signal.

  An interesting property of this signal is that the overall mean 
$r$ = $\Delta B$/$\Delta V$ ratio from analysis of the residuals is 
1.02 ${\pm 0.16}$. This value is indistinguishable from unity and is also 
consistent with the results of SHV. This confirms that the modulation 
originates in a region having the same color as the surface of the Be star.

  We take the following as our best ephemeris for the rotation signal:

\begin{equation}
T_{min} = RJD\,53000.407{\pm 0.024} + (1.215811{\pm 0.000030}) \times E.
\end{equation}

\noindent where the epoch,  referenced to light minimum
and period are mean values computed from the 
several $B$ and $V$ determinations of these quantities above, and the 
uncertainties are the standard deviations of the means.  Our new period 
makes only a slight improvement to $P = 1.21581 \pm 0.00004$ days from 
SHV, probably because the amplitudes were significantly lower in our new
observations.  

\begin{figure}[h!]
\centerline{
\includegraphics[scale=.35,angle=90]
{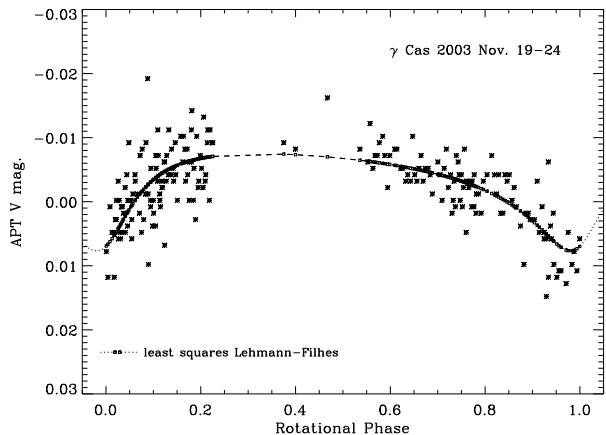}}
\caption{
Light curve of the $V$ residuals for 2003 November 19-24 
(RJD 52962--52967) phased to a period of 1.21581 days. Symbols
refer to observations taken on different nights during this interval.
The solid line is the fit from a least-squares Lehmann-Filhes solution 
($e$ = 0.49, $\omega$ = 40$^{\circ}$).
}
\label{nov03}
\end{figure}

  To show the brightness variability in detail, we plot in 
Figure\,\ref{nov03} the $V$ data residuals from the 
long cycles for six successive nights from 2003 November 19--24 against 
the mean  period.  This short subset is ideal for 
analysis because the long cycle has all but damped out during this period.  
Moreover, the data coverage on these nights is the densest we obtained for 
any consecutive nights in the entire program. Light curves for both 
passbands are exhibited in SHV.  Figure\,\ref{nov03} confirms the 
nonsinusoidal waveform found by SHV.  Its shape is reminiscent of an 
asymmetric radial-velocity curve of a binary system.  Therefore, we fit the 
data with a Lehmann-Filhes-type solution of the form:

\begin{equation}
m\,\,=\,\,K\,cos(\phi + \omega) + e\,cos( \omega),  
\end{equation}

\noindent where we used the period and time of a fiducial phase to 
determine the rotation phase $\phi$ and a semiamplitude K = 1.43\,mmag
from procedures described just below. This left three free parameters 
that were needed to model the waveform, a semiamplitude of 
0.0076$\pm{0.0010}$ mmag, a pseudo ``eccentricity" $e$ = 0.51$\pm{0.05}$, 
and a ``longitude of periastron" $\omega$ = 25$^{\circ}$${\pm 6}^{\circ}$; 
the full amplitude of course is 2K. Similar values were found for the $B$-band
dataset.  The formal {\it rms} is 0.0040, which is in the range of 0.0035-0.004 
we had already found for the nightly errors (SHV). Using this value in our 
Lehman-Filhes fitting gives a reduced $\chi^2$ statistic is 1.01 (206 degrees 
of freedom), which is essentially a consistency check on the input {\it rms} 
value. The corresponding reduced $\chi^2$ for a sinusoid of the same amplitude 
is 1.34.  As an independent check of our values, we rereduced the data for 
this figure by using the computer program BISP written by Dr. Frank Fekel 
that uses a modified form of the Wilsing-Russell method (Wolfe, Horak, \& 
Storer 1967), followed by a differential corrections program, SB1, by Barker, 
Evans, \& Laing (1967).  This effort resulted in an eccentricity
of $e$ = 0.53$\pm{0.05}$ and an $\omega$ = 6$^{\circ}$${\pm 6}^{\circ}$. The
analysis of the $B$ band data gave similar results.  
Note that together the $e$ and $\omega$ represent ``bowness" 
and ``asymmetry" attributes of the waveform. They are not independent and 
hence orthogonal parameters. For $e$ and $\omega$, the quoted errors were 
determined from the different values of these parameters from the 
individual $V$ and $B$ band light curves.  
Also, the observational errors for the $V$ and $B$ band datasets 
are likely highly correlated, so this is still not a satisfactory test for 
significance.  

  To address this issue of the statistical significance that the 
waveform is nonsinusoidal, we investigated the chances that a sine curve 
with the given noise properties could give a value of $e$ similar to 
0.53 that we found.  We subjected this result to a more
formal yet still conservative test by checking on the significance of the
``eccentricity" implied in the data and ignoring the asymmetry of the 
light curve, that is the cos($\omega$) component of the e\,cos($\omega$) 
term.  We began by forcing the solution of our data to a sine curve, that
is by adopting the previous period, mean light level, and amplitude
and running the Lehman-Filhes program once again.  This gave a somewhat
higher reduced chi-squares value of 1.34. We then ran simulations on the
best sine curve fit with the {\it rms} value we had found in our original
solution, allowing the $e$ parameter to be floating once again. We found
that it took 1.7 million realizations for the $e$ value to reach our
measured value of 0.53, suggesting that the nonsinusoidal eccentricity
parameter is significant at a value of 5$\sigma$.  Running the test once
again with a new initial seed value for the random number generator gave
almost the identical results. Meanwhile, we found that the lower bounds 
of our computed mean $\chi^2$ values in our simulations attained the value 
we found from our full-parameter Lehman-Filhes solution of our data at a 
slightly lower significance level of 4$\sigma$.
In view of these tests we conclude that the waveform was significantly
different from a sinusoid, and also shaded in a different direction from 
the right-shaded asymmetric waveform solved for by SHV over the 1997-2004
seasons.  
These conclusions are also consistent with the annual meandering of the fitted
phase of light minimum (again to a significance of several times our formal 
errors that result from our analysis of the changes in light amplitude 
described above.

\subsection{A search for other periods}

\begin{figure}[h!]
\centerline{
\includegraphics[scale=.35,angle=-90]
{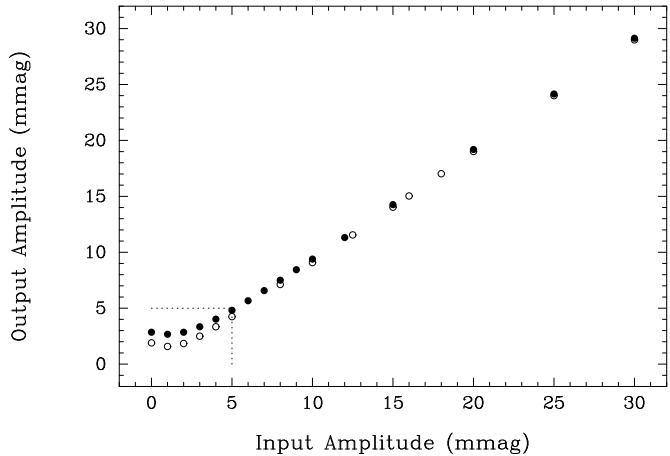}}
\caption{
The relation between artificial input signals at the orbital frequency
in our noisy light curves and their recoveries by periodogram
analysis. The recovery is linear and hence reliable to a level of
$\approx$5\,mmags. (faint dotted lines).
}
\label{recvr}
\end{figure}

  Finally, we searched for the influence of the secondary star on the light
curve through its known orbital period of 203.5 days.  Our main concern 
initially was to mitigate the effects of artificial low frequencies, 
principally in the seasonal window. We found that the effect of the lunar 
synodic period on the local sky was insignificant. Other studies
using the T3 APT have found the same null result.
The procedure that worked best was to ``condition" the 
data first by removing obvious (3$\sigma$) outliers and also 
removing the first observing season due to its zeropoint uncertainty.
The resulting ``raw" Fourier periodogram showed a dominant broad period 
centered at 351\,days. We attribute this to the interseason gaps associated 
with the Earth's orbital motion. A clutter of sidelobes due to the 
combinational frequencies of this and the diurnal frequency are present 
below 0.04 cy\,d$^{-1}$.  Most of the signal from this noise was mitigated 
by detrending of the 13 seasons of data.  The remaining (mainly combinational 
frequency) lobes were equivalent to signals at a level of 
$\approx$5-10\,mmags.  We inspected the resulting periodogram and found 
a 3\,mmag signal due to the rotational modulation discussed above. The low 
frequency range was dominated by a few weak signals associated with an 80\,day 
period (from Seasons 2003 \& 2007) and other incompletely removed cycles.  
However, we found no signal whatsoever at 0.0050 cy\,d$^{-1}$, associated 
with 203.5\,days or at any of its precise harmonics.  We then injected 
a series of sinusoidal signals of diminishing amplitudes of this frequency 
and attempted to recover them through their periodograms.  We discovered that
we could recognize an orbital frequency signature down to 2.5\,mmag (about 
the level of the rotational signal).  However, we were unable to recover 
the amplitudes below 5\,mmags.  Figure\,\ref{recvr} shows the results of
these simulations. Below 5\,mmags., the recovery becomes nonlinear and
unpredictable.  We take this value as the 2 sigma limit against the 
detection of a full-amplitude, orbitally-induced signal.

\section{Discussion}

\subsection{The outburst, long cycles and dynamos}

  Several important results have emerged from this study, and we start by 
discussing the long cycles, which were the first to be discovered by APT
photometry.  First, we have 
found no season in which the light curve does not betray some semblance 
of a cycle.  Second, although these oscillations are ubiquitous, they are 
not stable.  More often than not, the cycle lengths vary from year to year.  
As near as we can determine, the period continuity and phasing are unaffected 
when a new cycle is triggered and grows.  For example, we remind the reader 
of the peculiar event in the 2003 cycle in which an 80 day cycle continued 
at first from the previous year, quickly damped out, and grew again to a 
new 80-day cycle with the same phasing (SHV).  Thus, it may be more 
accurate to describe the cycles as originating from an excitation event 
in the disk, which damps out within one or two cycles.  As it weakens, the 
process begins anew as another cycle is excited on a timescale of as short
as two weeks (SHV).

  Third, we noticed in $\S$\ref{lng} that the fit for Season 2010 is poor 
(Fig.\,\ref{seas0813}).  This is just when the mass input from the
star was extraordinarily high, according to the X-ray results of S12.
We believe that the fit to the long cycles is flawed at this time
because of a sudden and temporary change in this loss rate rather than an
interruption in the disk cycle per se.  A temporary reduction of the
rate of the ongoing mass loss to the disk could also be responsible for the 
decreased brightness during the middle of Season 2005 as well. This 
conjecture is supported by the absence of the effect of mass-loss events 
on the cycle length or magnitude range ratio at other times. Only the 
rate of brightening is affected.

  The S12 study was indeed fortunate to observe \gam\ spectroscopically with 
{\it XMM-Newton} just as the star's outburst began.  This study recorded the 
effect of the mass loss incident on one of two X-ray absorption columns. 
This column covered ${\frac 14}$ of the X-ray emitting sources and doubled 
in 44 days.  (The second column remained constant.) 
Furthermore, during this time the column attained a historical high value of 
7.4$\times$10$^{23}$ atoms\,cm$^{-2}$, which was some 300 times higher than 
the historical minimum column length observed.  The significance is that the 
X-ray emission sites fall along the column to the source of this absorption, 
the Be star.  We have already noted that the concurrent cycle did not change 
except for the waning and waxing of its amplitude during the 2010 and 2011
seasons.  This is strong additional evidence that the long optical cycles 
arise in the disk.

  An additional marginally significant characteristic of the cycles 
is their potential dichotomy into two distinct populations that are defined 
by the gap in the $\Delta$B/$\Delta$V histogram (Fig.\,\ref{histr}). 
This implied bimodality could be important because it shows that the 
variations in the shifts in mean color $r$ convey physical information 
about the origin of the long cycles.  
Our interpretation of this possibility is that the cycle is 
excited in at least two somewhat different volumes. When they first suggested 
that the cycles originate within the disk, RSH could not be sure of even this
interpretation. The shifts in color are a strong indication indeed that
long cycles do not arise from the star's surface. We should also remark
that earlier concepts of the temperature stratification of disks assumed 
that these regions would be segregated simply by disk radius. However, 
recent models of Be disks (e.g., Sigut, McGill, \& Jones 2009) suggest 
that the denser regions of the disk cool from recombination processes. 
Since this cooling is proportional to electron density squared, a 
cooler emitting component in the disk plane is differentiated 
from ``warm sheaths" at some distance from the central plane.
A bimodality in the color ratio may be caused by a preference of cycles 
being excited either in the disk plane or at higher latitudes or, 
alternatively, at smaller versus larger disk radii.

  RSH and SHV speculated that the cycles arise from a disk dynamo, 
specifically via the Magnetorotational Instability (``MRI"; Balbus \& 
Hawley 1991, 1998). This mechanism is predicted to operate in any stellar 
disk isolated from external influences as long as an internal seed magnetic 
field exists and it undergoes Keplerian rotation.  The mechanism operates 
by a positive feedback loop relying on the existence of a random perturbation 
that forces disk matter to move radially and exchange angular momentum with
its surroundings.  The angular velocity shearing and the seed field excites a 
local turbulence.  These velocities amplify the initial seed field, causing 
an reamplification of turbulence, and so on. In the RSH picture, the positive 
feedback loop creates a dynamo in which disk zones alternately undergo matter 
compression and rarefactions during the cycle. The more highly compressed 
regions have higher electron densities and therefore higher emission measures, 
which is measured as light variations in the $V$ and $B$ passbands.  
For this MRI mechanism to be observable it must be unstable {\it globally.} 
In general, organization of modulating MRI cells cannot be predicted.  
Furthermore, if the magnetic field from the star intrudes into the disk, it 
will quench the mechanism.  We note also that simple arguments suggest that 
the cycle length should be of the order of the orbital period of particles in 
the disk -- a few days.  Nonetheless, the disparity between this value and 
the observed lengths, 2--3 months, is itself no disqualification since, for 
example, even the length of the solar 22-year cycle has yet to be predicted.

  Putting aside for a moment these obvious issues, the existence of a 
mechanism that segregates different constituents of the gas disk offers 
the possibility of changes from time to time in those regions where the 
MRI operates most efficiently.  To take one example, Emmanuel \& Balbus 
(2012) suggest that diffusion of dust grains toward the central plane of 
protoplanetary disks can create a ``dead zone" that suppresses the MRI.  
We cannot suggest the creation of grains in a Be star's disk.  However, 
one can imagine more realistic scenarios.  Perhaps higher than usual 
turbulent velocities in the less ionized central plane, where the gas is 
already more neutral, can decouple gas from the local field. 
Then, by limiting the MRI to 
operate in warmer (sheath) volumes only where the disk's contribution to 
the $B$ flux is greatest, the brightness range ratio $\Delta$$B$/$\Delta$$V$ 
would increase during the cycle. Such reasons could in principle give a 
range in the values of this color ratio over the a number of cycles.
Clearly this subject is highly speculative.

\subsection{The rotational modulation feature}
\label{rotmod}

   In this paper we have again recovered the 1.21 day period found by 
SHV and confirmed that it is robust over the full data span. 
Also, it undergoes no color changes during the period.
SHV presented reasons why the feature is rotational, among them that
the expected stellar radius (from HR Diagram and interferometry) and 
measured values of $v$sin\,$i$ and sin\,$i$ (interferometry)
predict a period of 1.25${\pm 0.17}$ days. At the time SHV analyzed their
results, both empirical results and models suggested that it seemed
unlikely that B0.5\,V stars would excite NRP modes with periods near 1 day. 
In the intervening years the launch of the CoRoT satellite (Baglin et al.
2006) and the bonanza of data it generated in fact has led to the 
discovery of NRP $g$ mode periods near 1 day in O9-B1.5 III-V stars 
(Degroote et al. 2009, Papics et al. 2011, Briquet et al.  2011), 
even though the excitation mechanism and the predicted properties
of these modes are still unclear (possibly because current internal 
opacity models are still inadequate). When the data sampling intervals 
are long and the signal to noise ratios are high enough, several modes 
are invariably discovered with periods from the $\beta$\,Cephei star range
of a few hours up to about 1.5 days.  Analysis of periodogram features
in the light curve of HD\,180642 indicates that the amplitudes of some 
modes change, while others are stable over timescales of up to three 
years (Degroote 2009). 
However, these stars differ from $\gamma$\,Cas by being slow to 
intermediate rotators, whereas $\gamma$\,Cas's rotation rate is almost 
indistinguishable from the critical velocity.

  In fact, despite the new work, no NRP modes with periods near an expected
rotational period have been found in these stars. This is not surprising
because most of the periods in the corotating frame of the star would then be
very long -- indeed, within the errors of their determinations, nearly infinite.
 However, it is possible in principle 
for very long periods to be observed if a  mode exists with a period that is 
long, but still somewhat shorter in the nonrotating frame, {\it and} if it 
is accompanied by a retrograde ($m > 0$) with a large rotational splitting. 
In principle, such splitting could give a frequency mode component that 
compensates for the positive frequency of its unsplit component mode 
($m$ = 0), again in the nonrotating frame. In this scenario only this one
retrograde mode would be detected.  While such a thought experiment is 
possible, we regard it as {\it ad hoc} and not very likely. 
Moreover, we have seen that this periodicity exhibits no color term
and no measureable phase difference between the $B$ and $V$ fluxes.
In contrast, NRP $g$ modes of high degree and low frequencies in the 
corotating frame have long been predicted and are well known to exhibit 
color changes over their cycles.
This is contrary to what we observe for $\gamma$\,Cas. Based on this fact
and on the unlikelihood of discovering a mode whose frequency in
the initial frame must be very close to the rotational frequency, we
reject the NRP hypothesis as the cause of the photometric 1.21 day signal.
Finally, as listed just below, there are independent reasons
for believing that disorganized surface magnetic fields exist on this star,
and these could lead to this signature.  

  The most important new results in this paper are that the amplitude of
the surface modulation decreased from 2003 to 2005, and the feature has been 
barely detectable ever since.  The occurrence of the 2010 outburst showed 
no correlation with the properties of this feature.  We have also discovered 
that the ``faint star" phase can vary significantly, as it did in 2003.  
This circumstance was traced to a change in the asymmetry of the waveform.  
Then, contrary to SHV's impression that rapid optical variations are 
influenced by simultaneous changes in local X-ray flux over several hours, 
we now see that the optical brightness variations on this and other nights 
are due mainly to rotational modulation.  From the present ephemeris, 
we find that the two major UV continuum dips and X-ray excesses observed 
by the {\it HST} and {\it RXTE} occurred at phases 0.45 (lesser UVC dip) 
and 0.78 (deeper dip).  According to the asymmetric waveform for this 
epoch (SHV), these phases correspond to a point on the brightening branch 
of the light curve and another just after brightness maximum.  

  Pertinent to our rotational analysis of \gam\ is the detection of 
rotational modulation signatures in other hot stars for which magnetic 
fields are not known to exist, either because they are below the level of 
detection or have not been searched for. In recent years, rotational 
signals have been found in light curves of HgMn stars such as 
$\alpha$\,And and AR\,Aur (e.g., Wade et al. 2006, Folsom et al. 2010).  
Analysis of Doppler imaging results for a number of HgMn stars, which are 
slow rotators ($v$sin\,$i$ $\approx$ 70 km\,s$^{-1}$), indicates that various 
elements on the surfaces of these stars are distributed inhomogeneously, 
that is, in discrete spots. Spectropolarimetric observations of 36 of these 
stars by Makaganiuk et al. (2011) have revealed no attendant magnetic 
field signatures.  These authors advance field/gas energy partition to
argue that magnetic field configurations in these spots cannot
hide these fields from detection and therefore that the spots have a 
nonmagnetic origin.  Rather, they speculate that spots arise from a yet 
undiscovered hydrodynamic instability that augments diffusion timescales 
locally by several orders of magnitude in a nonmagnetic environment.  
However, in the present application it appears that no argument can be 
reasonably advanced that these hypothetical effects create surface 
structures on more massive and rapidly rotating stars.

  Two other recent studies have found rotational modulation in main 
sequence B stars.  First, Papics et al. (2001) have discussed 
variability of light curves of HD\,51756 obtained by the CoRoT satellite.  
This is a double-lined binary system of which the primary is a B0.5\,IV 
star ($v$sin\,$i$ = 170 km s$^{-1}$) and the secondary is an early-type star
($v$sin$\,i$ = 28 km$^{-1}$ s$^{-1}$).  The light curve displays millimag level 
variabilities.  According to these authors, the variations must be due to 
harmonic structure of a single frequency, probably rotational modulation, and 
not to nonradial pulsations.  The variations can then be understood as being 
caused by a single surface disturbance, with $i$ $\approx$ 41.6$^{\circ}$  
if they arise from the primary or with $i$ $\sim$ 6.3$^{\circ}$ if they 
come from the secondary.  
Because magnetic measurements of this star have not yet been attempted, 
Papics et al. were not able to discuss the physical origin of the rotational 
modulation.  Interestingly, if it should be determined that the source of 
the modulation is on the primary, it would require the star to be rotating 
at near the critical velocity.  A reference to the Rossi All Sky Survey 
shows no strong X-ray source that could be identified with this system.  

  A second study showing rotational modulation in a normal abundance B star 
is that of McNamara, Jackiewicz, \& McKeever (2012). These authors obtained 
precise light curves of the star KIC\,005473826 (a likely late-type B star) 
using Kepler spacecraft data and found two frequencies. The  primary signal 
occurs at 0.95 d$^{-1}$. A second frequency twice this value is likely to be 
its harmonic.  The amplitudes of these two frequencies are 10.8 mmags and 
6.0 mmags, respectively, i.e., they are larger than we have found for 
$\gamma$\,Cas. These authors ascribe the signal to rotational modulation 
of a magnetic spot.  This star seems to be beyond the reach of present 
spectropolarimetric systems for a direct magnetic detection.

  We conclude this discussion with a summary of arguments
for surface magnetic fields on the surface of the Be star \gam\, 
{\it apart from the rotational surface feature}.  Each of the following is 
an independent argument for the existence of matter being contained in 
a volume situated above one or more regions of the Be star's surface:

\begin{enumerate}
\item the existence of the {\it msf} 
moving rapidly across optical and UV line profiles in the star's spectrum. 
This implies the corotation of small, short-lived cloudlets (Smith, Robinson, 
\& Hatzes 1998, ``SRH"; Smith \& Robinson 1999, ``SR"). 

\item the above mentioned brief dips over several hours in the UV continuum, 
indicating the existence of 3--4 optically thin, corotating clouds. Other
dips have been found in light curves formed from contemporaneous {\it IUE} 
data (SRH). Surface starspot models can be ruled out from their rapid 
evolution of the transits in time and from the redness of their color in 
the UV wavelength domain (SRH).

\item the existence of corotating clouds having unexpected temperatures 
near the surface of a B0.5 star. 
Simultaneous observations suggest that the ionization of Fe and Si in 
regions over the star responded to changes in X-ray flux (Smith \& Robinson 
2003).  
\end{enumerate}

  Each of these arguments supports the presence of circumstellar matter
close to but not on the stellar surface, unlike the rotational feature.

  The evidence for magnetic fields being responsible for the spots is quite
strong.  The fact that they are likely to be disordered and to have decreased 
in the last few years at least partially addresses the concern of Neiner et 
al. (2012) that a spectropolarimetric observation during this period did not
lead to a positive result.  Depending on the complexity of their multipolar 
structure, it can be hoped that future observations can discover them when 
the amplitude of the photometric modulation recovers to its its former value.  
It is important to point out in any case that 
such fields must necessarily have complex toplogies, since they would 
otherwise be easily inferred from Bp signatures (cyclical emission/absorptions 
in the UV resonance lines of Si, C, and N) that accompany dipolar fields.

\section{Conclusions}
\label{concl}

  For over two decades, the overriding question concerning \gam\ has been 
the mechanism behind the production of its hard X-rays.  Ironically, it is 
possible that the ultimate resolution of this question will not come so 
much from continued X-ray observations of the \gam\ stars as much as
traditional optical studies. First, radial velocities must be derived to 
determine whether binarity is a necessity for the $\gamma$\,Cas-like X-ray 
properties to develop, and, if so, what the orbital separations and 
eccentricities are -- such parameters are relevant to the determination 
of the X-ray flux generated by accretion of mass by the secondary.  Their 
role in binaries may also determine why they reside in such a small region 
of the HR Diagram and, indeed, whether this determines a particular type 
of evolved secondary.

  Second, and pertinent to this study, is the role of traditional optical 
light monitoring.  Despite attempts to start such work on the next brightest 
star, HD\,110432, such work has not yet taken root in the Southern hemisphere.
Even so, there is a great deal of work that could be done to characterize 
its optical variations.  If this star also shows rotational variability, 
the $v$sin\,$i$ suggests the rotational period is also near one day.   
This is clearly a task best suited to an APT or space-borne telescope.

  Our \gam\ campaign has led to several unanticipated discoveries pertinent 
to the X-ray excitation mechanism. These include: 

\begin{enumerate}
\item long optical cycles, probably often correlated with XR cycles (RSH, SHV), 
\item optical brightenings/reddening in the disk correlated with attenuation
      of XR cycles,
\item partial damping of long cycle amplitudes due to the mass loss events,
\item cycles exciting in somewhat different regions of the disk,
\item rotational signal and waveform varies over time. 
\end{enumerate}

  Concerning this last point, the extended and asymmetric minimum appearing 
in the SHV solution of the average for Seasons 1997--2004 probably arises from 
a superposition of changing waveforms during this period.  This means that 
whereas the centroid of the surface feature that causes this is stable, 
the internal distribution of the disturbing property, assumed to be due 
to magnetic fields, is not. Moreover, the ``sharpness" or even the
``asymmetry" cannot be considered a permanent and defining characteristic
of the light curve.

  We noted in $\S$\ref{rotmod} that the UV continuum dips and X-ray excesses 
observed by SRH coincide with APT rotational phases 0.45 and 0.78, i.e., 
shortly before brightness maximum and near the brightness null.  
This phasing suggests one of two possible explanations.

  The first is the mechanism producing rotational modulation in 
this B0.5e star is fundamentally different from those of later B-type 
stars -- that is, a magnetic structure causes a local optical 
{\it brightening}.  Support for this view comes from the discovery in
visible-band light curves of two magnetic Of?p stars, HD\,191612 and HD\,108,
that maximum brightness coincides with the transits of magnetic 
spots and maximum H$\alpha$ emission (Holwarth et al. 2009; Barannikov 2007; 
Martins et al.  2010).  A theoretical construct for bright magnetic spots 
has been advanced by Cantiello \& Braithwaite (2011).  Their argument is 
that magnetic pressure in the spots partially offsets gas pressure once 
pressure equilibrium is established, and this reduces the radiative opacity 
along a column.  The lower opacity allows photons to escape from a deeper, 
hotter and, therefore, brighter photospheric region than would be the case 
in a nonmagnetic atmosphere.  
(This is essentially the reason that faculae appear bright 
above thin magnetic tubes on the solar surface.)  According to these authors, 
these arguments should apply for stellar masses down to about 10\,M$_{\odot}$. 
However, for the magnetic Bp stars (i.e., for masses up to 9\,M$_{\odot}$, 
or type B2),  it is {\it known} that the optical flux decreases during the 
transits of magnetic spots.  Then for these stars the physical explanation must 
be altogether different.  In Bp star light curves, 
periodic decreases in optical flux are caused 
by the appearances in the optical continuum of bound-free absorption edges 
caused by superabundant heavy elements as chemically peculiar spots 
(approximately coinciding with magnetic spots) transit the star's surface.  
Given an assumed mass of 14-15\,M$_{\odot}$ (e.g., Harmanec 2002; Gies et 
al. 2007), it seems on the face of it that the explanation for the 
magnetic O star variability could also hold for $\gamma$\,Cas.\footnote{SHV 
attempted to tie the rotational signature of \gam\ to the phenomenology of 
other magnetic Bp stars like HD\,37776.  The current discussion 
should correct their speculation on this point.}

  A second, perhaps equally compelling, interpretation is that the inferred,
active X-ray center/cloud complexes observed in 1996 January in \gam\ do 
not reside on fixed locations at all.  In this 
explanation the clouds producing those UV dips were transient and 
occurred at no special longitudes on the Be star's surface. This carries 
the important implication that the circumstellar clouds causing them and
therefore the underlying X-ray active structures, do not remain at fixed
longitudes. In fact, in the RS scenario X-rays are produced by 
surface impacts of high energy electron beams originating from external 
magnetic field relaxations.  If these beams are not channeled to preferred 
surface locations by open field lines anchored to permanent spots, 
the ensuing flare events might well occur at essentially random areas
on the Be star. In this case, the current bright-star
alignment with the X-ray active spots in 1996 was coincidental.

  Going forward, we emphasize that studies of \gam\ and other analogs must not 
only define their properties as a group but even more importantly, {\it they
must also determine what sets them apart from other B0.5--B1e stars that rotate 
very rapidly and yet do not show abnormal X-ray emissions.}  In addition, 
having been building now for several decades, perhaps the disk of \gam\
will start to decay within a few years.  It will then become critical to find
out whether the emission continues when the disk decays but is still strong. 
In this circumstance, the disk could no longer plausibly transfer mass 
through resonance interactions with its companion, and the last plausible 
efficient conduit of matter transfer to a degenerate companion will be lost.  
This will provide an important test of whether X-rays are generated via 
blob accretion onto a white dwarf.  Thus, the coming goals of a continuing 
APT monitoring program will be to await both the end of the contemporary 
disk building era and (independently) the restoration of the rotational 
modulation to its former strength. The latter event will offer a better chance 
of discovery of a magnetic field by current spectropolarimetric techniques.

  The authors would like to express their appreciation to the referee for
valuable scientific comments that improved this paper. We also profited from 
the use of two orbital-velocity solution codes written by Dr. Frank Fekel.
GWH acknowledges long-term support from NASA, NSF, Tennessee State University, 
and the State of Tennessee through its Centers of Excellence program. MAS
acknowldedges funding of NASA Grant NNX11AF71G through the
to Catholic University of America under the Advanced Data Analysis Program.

\end{document}